\documentclass[showpacs,preprintnumbers,amsmath,amssymb,aps, prd]{revtex4}
\usepackage{graphicx}
\usepackage{grffile}
\usepackage{mathrsfs, mathtools}
\usepackage{caption}
\usepackage{float}
\usepackage{xcolor}
\usepackage{changepage}
\usepackage{dcolumn}
\usepackage{bm}
\usepackage{graphicx,subfigure,epsfig}
\usepackage{multirow}

\def\e{\eta}
\def\x{\xi}
\def\c{\chi}
\def\tep{\tilde{\epsilon}_v}
\def\tet{\tilde{\eta}_v}
\def\p{\phi}
\def\n{{\cal N}}
\def\mp{M_{pl}}
\def\l{\lambda}
\def\a{\alpha}
\def\b{\beta}
\newcommand\be{\begin{equation}}
\newcommand\ee{\end{equation}}
\newcommand\ba{\begin{eqnarray}}
\newcommand\ea{\end{eqnarray}}
\newcommand\bt{\bibitem}

\newcommand\lt{\left}
\newcommand\rt{\right}
\begin{document}
\title{Non-Minimal $RT$ Coupling and its Impact on Inflationary Evolution in $f(R,T)$
Gravity}
\author{Sohan Kumar Jha}
\affiliation{Chandernagore College, Chandernagore, Hooghly, West
Bengal, India}
\email{sohan00slg@gmail.com}
\author{Anisur Rahaman}
\email{anisur.rahman@saha.ac.in;
manisurn@gmail.com (Corresponding Author)}
\affiliation{Hooghly Mohsin College, Chinsurah, Hooghly - 712101,
West Bengal, India}

\date{\today}
\begin{abstract}
\begin{center}
Abstract
\end{center}
We examine inflationary models in the $f(R, T)$ gravity framework where we have a conformal constant and an $RT$-mixing term apart from an R term. The RT-mixing term introduces non-minimal coupling between gravity and matter. We consider the exponential SUSY potential $V(\p)=M^4 \lt(1-e^{-\l \p/\mp}\rt)$ and a novel potential $V(\p)=\l \mp^{4-2\a} \p^{2\a} \sin^2\lt(\frac{\b \mp^\a}{\p^\a}\rt)$. With the help of COBE normalization, we constrain values of different parameters and extract the field value at the time of Hubble crossing. The end of inflation is marked by $\tep(\p_i)=1$ where $\p_i$ is the field value at the end of inflation. Equipped with these values, we then move on to calculate values of spectral index $n_s$ and tensor-to-scalar ratio $r$. Our predicted values of $n_s$ and $r$ fall within their observed values from the Planck 2018 survey and BICEP/Keck array measurement for both potential, making them plausible candidates for the inflationary model. We also display the variation of the tensor-to-scalar ratio and spectral index with the coefficient of RT-mixing term for fixed values of e-fold number. There, we find the existence of two local maxima of $n_s$, which occur at a negative and a positive value of $\x$, the coefficient of $RT$-mixing term. Our analysis finds a significant impact of $\x$ on values of observables.
\end{abstract}
\maketitle
\vspace*{0.5cm}

Kkeywords: Cosmology, Inflation, Modified gravity, Planck, BICEP/Keck, RT-mixing

\section{Introduction}
The history of the observable Universe prior to the era of nucleosynthesis remains largely unknown. However, it is widely accepted that cosmic inflation - a phase of exponentially rapid expansion occurred shortly after the Big Bang. This theoretical framework, proposed initially by Alan Guth \cite{GUTH}, offers solutions to several problems related to cosmology, notably the horizon and flatness issues. This theory posits a brief but extremely rapid expansion of the Universe, thereby smoothing out initial irregularities and establishing a nearly uniform temperature across vast distances. The concept of inflation was further refined by Linde \cite{LINDE} and Starobinsky \cite{STARO}, becoming a foundational element of modern cosmology.

Over the past few decades, our understanding of cosmic dynamics has advanced considerably, fueled by the developments of theoretical modeling and observational techniques. The $\Lambda$-Cold Dark Matter ($\Lambda$CDM) theory has become the prevailing framework in modern cosmology, operating within the General Relativity (GR) scope. It effectively explains the formation of large-scale structures as well as the accelerated expansion of the Universe. The $\Lambda$CDM model is supported by high-precision measurements of the Cosmic Microwave Background (CMB), particularly from COBE \cite{COBE}, WMAP \cite{WAMP}, and Planck \cite{PLANCK}, which provide crucial insights into the temperature, density, and large-scale features of the early Universe. Despite its many successes, some challenges remain unresolved within this model.

Inflation theory resolves flatness and horizon issues by proposing that the Universe experienced exponential expansion in its earliest moments, effectively homogenizing and flattening the observable Universe. In addition, inflation provides an excellent avenue for generating the small density fluctuations that later evolved into the large-scale structures observed today. These fluctuations are well described by slow-roll inflation models, which assume that the kinetic energy of inflation is negligible relative to its potential. Observational data from WMAP \cite{WAMP} and Planck \cite{PLANCK} strongly support the predictions of such models, particularly the nearly scale-invariant power spectrum of density fluctuations.

Although inflationary models, especially those involving scalar fields and specific potentials, are widely accepted, many questions remain about the very early Universe, particularly the period preceding nucleosynthesis \cite{LYTH}. This has led to growing interest in modified gravity theories that extend beyond GR and $\Lambda$CDM, aiming to explain phenomena such as dark energy, dark matter, and the late-time acceleration of the Universe. Among these,
$f(R)$ gravity \cite{HARKO1, ROSHAN},
$f(R,L_m)$ gravity \cite{NOJI1, ALLEM, SAHOO, BISABR}, and
$f(\phi,T$) gravity \cite{CHEN, GODA} have garnered significant attention. In particular, the $f(R, T)$ gravity model modifies the Einstein-Hilbert action by incorporating arbitrary functions of $R$ and $T$ where $R$ is the Ricci scalar ($R$), and $T$ is the trace of the energy-momentum tensor ($T$). This offers a richer structure for exploring cosmological evolution at large and local scales.

The  $f(R, T)$ gravity framework has been applied to a variety of cosmological contexts, including the study of dark matter and dark energy \cite{HARKO1}, thermodynamics \cite{HARKO2}, and scalar perturbations \cite{HARKO3}. It has also proven useful for investigating inflation, particularly in scenarios involving quadratic potentials and perfect fluids \cite{HARKO4, NOJI1, FRTQU, RTFLUID}. However, many of these studies rely on simplifying assumptions, such as minimal matter-gravity coupling, which may limit their generality \cite{HARKO4, NOJI1, FRTQU, RTFLUID, FRT10, FRT12, FRT13, FRT14}.

Our research aims to address this gap by presenting a more comprehensive study of inflation in
$f(R, T)$ gravity, with particular emphasis on non-minimal matter-gravity couplings. These couplings challenge the conventional view that matter is only weakly coupled to gravity. In such models, the energy-momentum tensor becomes non-conservative, which permits irreversible processes like particle creation - phenomena often investigated within the realm of quantum field theory in curved spacetime.

Current observational constraints have ruled out several inflationary models, including chaotic and natural inflation. To reconcile theoretical predictions with data, we introduce a linear term in
$T$, an $RT$ mixing term functioning as a rescaling mechanism, and the standard
$R$ term from Einstein's GR. The inclusion of the $RT$ mixing term facilitates the non-minimal matter-gravity couplings. This significantly impacts inflationary dynamics and enhances the predictive power of the model, allowing concordance with observational constraints on key inflationary parameters such as the number of e-folds and the tensor-to-scalar ratio $r$.

This work intends to widen the inflationary paradigm with the incorporation of more general conditions within the $f(R, T)$ gravity framework. This broadens the class of viable inflationary models and deepens our understanding of the Universe's early evolution. While inflation has successfully addressed core cosmological challenges like the horizon and flatness problems, further exploration of modified gravity theories remains essential. The $f(R, T)$ framework offers a potent tool for refining inflationary models and advancing our knowledge of cosmic evolution from the Big Bang to the present \cite{CAPO, ALVA, NOJI2, CHEN1, NOJI3, BASS, MARTIN, CHOW, FARONI1, FARONI2, VD1,VD2, LABA, GAMO, ALVES, FISHER, 
 BHATT, DEV}. The exponential SUSY potential $V(\p)=M^4 \lt(1-e^{-\l \p/\mp}\rt)$ \cite{susy1, susy2, susy3, susy4} and a novel potential $V(\p)=\l \mp^{4-2\a} \p^{2\a} \sin^2\lt(\frac{\b \mp^\a}{\p^\a}\rt)$ \cite{pot} have been considered here for a comprehensive and precise analysis of inflationary dynamics within the
$f(R, T)$ gravity framework, where the mixing term $RT$ is also explicitly taken into account. Both potentials are physically motivated and have appeared in cosmology and inflationary model-building contexts.
The exponential potential is often used in supersymmetric (SUSY) theories and string-inspired inflation models.
The second "novel" potential is an interesting hybrid form combining power-law and oscillatory behavior, which can lead to rich inflationary phenomenology. The analysis of inflationary dynamics using these two distinct potentials within the
$f(R, T)$ framework, particularly in the presence of the
$RT$ mixing term offers both a compelling and instructive avenue for investigation. The inclusion of the non-minimal coupling term $RT$
significantly modifies the gravitational sector and affects the evolution of the inflaton field and the dynamics of the early Universe. Such models deepen our understanding of inflation in modified gravity scenarios and enable us to test their viability against current observational constraints from Planck and other CMB experiments. Comparing the behavior of well-motivated potentials, such as the exponential SUSY potential and new phenomenological potential within this extended framework, could potentially reveal signatures unique to $f(R, T)$ gravity.

We organize the remainder of this as follows:
Section II is where we provide a concise overview of the field equations in modified gravity, incorporating the
$RT$ mixing term. A subsection within this section presents a general framework for inflationary dynamics, including the derivation of precise slow-roll conditions.
Section III is dedicated to computing the inflationary parameters for the exponential SUSY potential. It also presents a graphical comparison of the theoretical predictions with recent observational data. Section IV follows an approach similar to Section III, with a focus on the novel phenomenological potential.
Finally, Section V concisely discusses the obtained results and presents the study's overall conclusions.

\section{Field equations in modified gravity with RT mixing term}
The scenario where the inflaton field drives
inflation, the action of the the $f(R, T)$ type modified gravity is
provided by
\begin{equation}
\mathcal{S}=\int d^4x\sqrt{-g}\left[\frac{1}{2\kappa^2}f(R,T)+\mathcal{L}_m\right],
\label{ACTION}
\end{equation}
where $R$ is the  Ricci scalar, $T$ is the trace of
energy-momentum tensor $T_{\lambda\rho}$, and $g$ stands for the
determinant of metric tensor $g_{\lambda\rho}$. $G$ and  $\kappa$
have the definitions  $\kappa^2=8\pi,  G=\frac{1}{M_{Pl}^2}$
respectively. We set $\kappa$  and $G$ to unity during our
analysis.

We consider a precise algebraic form of $f(R, T)$ modified gravity
model, which is given by
\begin{equation}
f(R,T)=R \bigl(1+\eta+\kappa^4\xi T \bigr)+\kappa^2\chi T,
\label{FRT}
\end{equation}
where $\eta,~\xi$ and $\chi$ are the modified gravity parameters.
One can recover Einstein's gravity by taking the limit
$\eta,\xi,\chi$ $\rightarrow 0$. The energy-momentum tensor of the
scalar field  reads
\begin{equation}
T_{\lambda\rho}=\partial_{\lambda}\phi\partial_{\rho}\phi+g_{\lambda\rho}\mathcal{L}_m,
\end{equation}
where the Lagrangian density of the scalar field is given by
\begin{equation}
\mathcal{L}_m=-\frac{1}{2}g^{\lambda\rho}\partial_{\lambda}\phi\partial_{\rho}\phi-V(\phi).
\label{LSCA}
\end{equation}
The term   $V(\phi)$ in Eqn. (\ref{LSCA}) is representing the potential for
the scalar $\phi$.

By varying the action standing in Eqn. (\ref{ACTION}) with
respect to metric $g_{\lambda\rho}$, the field equations
corresponding to the modified model is obtained in the same
manner we do in computing the field equation for the action
of the standard Einstein's Gravity:
\begin{equation}
f_R R_{\lambda\rho}-\frac{1}{2}f
g_{\lambda\rho}+\bigl(g_{\lambda\rho}\square-\nabla_{\lambda}\nabla_{\rho}
\bigr)f_R=T_{\lambda\rho}-f_T
\bigl(T_{\lambda\rho}+\Theta_{\lambda\rho} \bigr). \label{FIELD}
\end{equation}
Here $f_R = \frac{\partial f}{\partial R}$, $f_T = \frac{\partial
f}{\partial T}$ and $\Box = \nabla_\lambda \nabla^\lambda$, where
$\nabla_\lambda$ is the covariant derivative. The tensor
$\Theta_{\lambda\rho}$ is defined in terms of the energy-momentum
tensor as
\begin{equation}
\Theta_{\lambda\rho}= g^{\lambda \kappa} \frac{\delta T_{\lambda \kappa}}{\delta g^{\lambda\rho}}
=-T_{\lambda\rho}-\partial_{\lambda}\phi\partial_{\rho}\phi.
\end{equation}
\subsection{Inflation: Slow-roll conditions}
In this section, we will list all of the equations required to compute inflationary
observables. When inflation occurs, the universe
is thought to be homogeneous and flat. So, the following Friedmann-Robertson-Walker (FRW) metric
provides a decent description of the spacetime:
\begin{equation}
ds^2=-dt^2+ a(t)^2[dx^2+dy^2+dz^2],
\end{equation}
where  $a(t)$ represents the scale factor. It effectively
represents the size of the universe, and by solving the field
equations of the theory, one may determine how the cosmos has
changed during cosmic time using Eqn. (\ref{ACTION}). Recasting the theory
into Einstein's frame through field redefinitions and conformal
transformations is a common technique that could be regarded as a
standard trick when considering inflationary models under modified
theories of gravity. One well-known example is the Starobinsky
inflationary model, which was developed from an action with a
$R^2$ term in the Jordan frame. The dynamical scalar degree of
freedom linked to the $R^2$ term was discovered to be represented
as a scalar field minimally connected to the Einstein-Hilbert
action when the theory was recast into the Einstein frame. The
Einstein's frame allows for a straightforward computation of the
inflationary observables. However, Ref. \cite{STARO} demonstrates
that Einstein's frame representation usually vanishes when non-minimally derivative couplings are incorporated into
the theory.
Therefore, the updated theory cannot display inflationary
observables using Einstein's frame representation when a non-zero
$\xi$ is present.

Calculating the inflationary observables by converting the system to Einstein's frame is still feasible as long as one ignores
higher-order derivative elements in the equations before
proceeding to Einstein's frame and consider the slow roll
approximations, which are a necessary conditions for the inflation
scenario of the early universe. According to slow-roll
inflationary models, the scalar field is rolling slowly on its
potential and all quantities, including $H$ and $\phi$, are
assumed to be evolving quasi-statically. The slow-roll
requirements for inflationary growth include
\begin{eqnarray*}
&&|\dot{\phi}|<< V(\phi), \nonumber \\
&&|\dddot{\phi}|<<|H\ddot{\phi}|<<|H^2\dot{\phi}|, \nonumber \\
&&|\ddot{H}|<<|H\dot{H}|<<|H^3|.
\end{eqnarray*}

Using the specific form of $f(R,T)$ given in equation (\ref{FRT}),
Einstein's equation takes the following modification
\begin{equation}\label{EINST}
R_{\lambda\rho}-\frac{1}{2}Rg_{\lambda\rho}=T_{\lambda\rho}^{E},
\end{equation}
where $T_{\lambda\rho}^{E}$ is the effective energy-momentum tensor.
It is provided by
\begin{equation}
T_{\lambda\rho}^{E}=T_{\lambda\rho}+\bigl(R \xi+\chi
\bigr)\partial_{\lambda}\phi\partial_{\rho}\phi-R_{\lambda\rho}
\bigl(\eta+\xi T \bigr)+\frac{1}{2}g_{\lambda\rho} \Bigl\{R
\bigl(\eta+\xi T \bigr) + \chi
T\Bigr\}-\bigl(g_{\lambda\rho}\square-\nabla_{\lambda}\nabla_{\rho}
\bigr)\xi T.
\label{TMN}
\end{equation}
The effective energy density and effective pressure resulted from the energy-momentum tensor
are given by the following expressions:
\begin{equation}
\rho^{e}=\frac{1}{2}\dot{\phi}^2 \Bigl(1+12\xi\dot{H}+\chi+18\xi H^2 \Bigr)
+V \Bigl(1+2\chi+12\xi H^2 \Bigr)-3\eta H^2-6H\xi\dot{\phi}\ddot{\phi}+12\xi H\dot{\phi}V_{,\phi},
\end{equation}
\begin{equation}
\begin{split}
p^{e}=&\frac{1}{2}\dot{\phi}^2 \Bigl(1+\chi+4\xi\dot{H}+6\xi H^2 \Bigr)-V \Bigl(1+2\chi+8\xi\dot{H}+12\xi H^2 \Bigr)+2\eta\dot{H}+3\eta H^2+\xi \Bigl(2\ddot{\phi}^2+2\dot{\phi}\dddot{\phi}\\
& -4\ddot{\phi}V_{,\phi}-4V_{,\phi\phi}\dot{\phi}^2+4H\dot{\phi}\ddot{\phi}-8H\dot{\phi}V_{,\phi} \Bigr).
\end{split}
\end{equation}
Eqn. (\ref{EINST}) with the use of  Eqn. (\ref{TMN}) leads to  the modified Friedmann equation
\begin{equation}
\label{FRIED1}
3H^2 \bigl(1+\eta \bigr)=\frac{1}{2}\dot{\phi}^2 \bigl(1+12\xi\dot{H}+\chi+18\xi H^2 \bigr)
+V \bigl(1+2\chi+12\xi H^2 \bigr),
\end{equation}
and consequently, the acceleration equation reads
\begin{equation}
\label{FRIED2}
\begin{split}
\frac{\ddot{a}}{a} \bigl(1+\eta \bigr)=
&-\frac{1}{3}\biggl\{\dot{\phi}^2\Bigl(1+\chi+6\xi\dot{H}-6\xi V_{,\phi\phi}+9\xi H^2 \Bigr)-V\Bigl(1+2\chi+12\xi\dot{H}+12\xi H^2\Bigr) \biggr\}\\
& \qquad -\xi H\dot{\phi}\ddot{\phi}+2\xi H\dot{\phi}V_{,\phi}-\xi\ddot{\phi}^2-\xi\dot{\phi}\dddot{\phi}+2\xi\ddot{\phi}V_{,\phi}.
\end{split}
\end{equation}
Additionally, the inflaton field satisfies the following equation of
continuity:
\begin{equation}
\label{EOM}
\ddot{\phi}\Bigl(1+\chi+6\xi\dot{H}+12\xi H^2\Bigr)
+V_{,\phi}\Bigl(1+2\chi+12\xi\dot{H}+24\xi H^2\Bigr)
+3H\dot{\phi}\Bigl(1+\chi+10\xi H^2+8\xi \dot{H}\Bigr)+6\xi\dot{\phi}\frac{\dddot{a}}{a}=0.
\end{equation}
In the limit $\eta \rightarrow 0$, $\xi \rightarrow 0$, and $\chi \rightarrow 0$, the above
set of equations subsequently turns into
\begin{equation}
3H^2 = \frac{1}{2}\dot{\phi}^2+V(\phi), \label{EQMA}
\end{equation}
\begin{equation}
\frac{\ddot{a}}{a} = -\frac{1}{3}(\dot{\phi}^2 - V ), \label{EQMB}
\end{equation}
\begin{equation}
\ddot{\phi}+3H\dot{\phi}+V_{,\phi}=0. \label{EQMC}
\end{equation}
The equations (\ref{EQMA}), (\ref{EQMB}), and {\ref{EQMC})
are respectively the  Friedmann equation, acceleration equation,
and  equation of continuity for the usual Einstein's gravity with
a minimally coupled scalar field.
Implementing these slow-roll conditions, Eqn. (\ref{FRIED1})
and Eqn. (\ref{FRIED2})  are approximated as follows:
\begin{equation}
H^2  \approx \frac{V}{3}\Biggl[\frac{1+2\chi}{1+\eta-4\xi
V}\Biggr], \label{HUBL}
\end{equation}
and
\begin{equation}
\dot{H}\approx -\frac{1}{2}\dot{\phi}^2\Biggl[\frac{(1+\eta)
(1+\chi)+4\xi V\chi}{(1+\eta-4\xi V)^2}\Biggr], \label{ACC}
\end{equation}
and the equation of continuity (Eq. \ref{EOM}) subsequently turns into
\begin{equation}
3H\dot{\phi}\Bigl\{(1+\chi)(1+\eta)+4\xi V\chi\Bigr\}
+V_{,\phi}(1+2\chi)(1+\eta+4\xi V)\approx 0.
\label{CONT}
\end{equation}

We are now in a position  to express   the above three equations in the
Einstein's frame. Three conformal transformations  are required in
this instance to obtain the desired  metric $g_{\lambda\rho}$, the
inflaton field $\phi$, and the inflationary potential $V(\phi)$ which will be
concomitant to Einstein's frame. The transformations are
$\Omega(\phi),~\tilde\Omega(\phi)$, and $\hat\Omega (\phi)$ respectively. The
first conformal transformation $\Omega(\phi)$  is associated with
the transformation of metric $g_{\lambda\rho}$ to an auxiliary
metric $\tilde{g}_{\lambda\rho}$:
\begin{equation}
\tilde{g}_{\lambda\rho}=\Omega(\phi)g_{\lambda\rho}.
\end{equation}
It leads us to write down the  line element  as
\begin{equation}
\tilde{g}_{\lambda\rho}dx^{\lambda}dx^{\rho}=-d\tilde{t}^2
+\tilde{a}(\tilde{t})^2[dx^2+dy^2+dz^2]
=\Omega(\phi)\left[-dt^2+a^2(t)(dx^2+dy^2+dz^2)\right].
\end{equation}
The second conformal transformation $\tilde\Omega(\phi)$ is
associated with the transforms of the field $\phi$ to an auxiliary
scalar field $\tilde{\phi}$. It is given by
\begin{equation}
\Biggl[\frac{d\tilde{\phi}}{d\phi}\Biggr]^2=\tilde\Omega(\phi),
\end{equation}
and  the final, i.e.,  the third transformation, is accompanied by
the transformation of the scalar potential $V(\phi)$. Under the
the conformal transformation   $\hat\Omega(\phi)$ the potential in
the Einstein's frame takes the following form $V(\phi)$
\begin{equation}
\tilde{V}(\tilde{\phi})=\hat\Omega(\phi)V(\phi).
\end{equation}
Applying these transformations on the metric, the scalar field,
and the scalar potential, we can write Eqns. (\ref{HUBL}),
(\ref{ACC}), and (\ref{CONT}) in the Einstein frame as
\begin{equation}
\tilde{H}^2\Omega(1+\eta-4\xi V)=\frac{V}{3}(1+2\chi).
\end{equation}
It can be approximated as
\begin{equation}
3\tilde{H}^2\approx \tilde{V}.
\end{equation}
Additionally, we have
\begin{equation}
\dot{\tilde{H}}\Omega(1+\eta-4\xi V)=-\frac{1}{2}\dot{\phi}^2
\Biggl[\frac{(1+\eta)(1+\chi) +4\xi V\chi}{1+\eta-4\xi V}\Biggr],
\end{equation}
which implies
\begin{equation}
\frac{d\tilde{H}}{d\tilde{t}}\approx
-\frac{1}{2}\Biggl[\frac{d\tilde{\phi}}{d\tilde{t}}\Biggr]^2,
\end{equation}
and
\begin{equation}
3\tilde{H}\frac{d\tilde{\phi}}{d\tilde{t}}+\tilde{V}_{,\tilde{\phi}}\approx 0,
\end{equation}
where $\tilde{V}_{,\tilde{\phi}}=\frac{d\tilde{V}}{d\tilde{\phi}}$,
and
$\tilde{H}=\frac{\dot{\tilde{a}}}{\tilde{a}}=H\Omega^{-\frac{1}{2}}$,
$\dot{H}=\Omega\dot{\tilde{H}},~\dot{\phi}=\frac{d\tilde{\phi}}{d\tilde{t}}\Omega^{-\frac{1}{2}}\tilde\Omega^{-\frac{1}{2}}$.
The above set of equations can be obtained by choosing
\begin{equation}
\Omega=1+\eta-4\xi V,
\end{equation}
\begin{equation}
\tilde\Omega=\frac{(1+\eta)(1+\chi)+4\xi\chi V}{(1+\eta-4\xi V)^2},
\end{equation}
and,
\begin{equation}
\hat\Omega=\frac{1+2\chi}{(1+\eta-4\xi V)^2}.
\end{equation}
In the Einstein's frame, the salient and potential slow roll parameters can be
defined as
\begin{eqnarray}\label{SRL}
\tilde{\epsilon}_v =
\frac{1}{2}\Biggl[\frac{\tilde{V}'(\tilde{\phi})}{\tilde{V}(\tilde{\phi})}\Biggr]^2,
~~\tilde{\eta}_v =
\frac{\tilde{V}^{''}(\tilde{\phi})}{\tilde{V}(\tilde{\phi})},
\end{eqnarray}
where \begin{equation*}
\tilde{V}'(\tilde{\phi})=\frac{d\tilde{V}}{d\tilde{\phi}}=\tilde\Omega^{-\frac{1}{2}}\frac{d}{d \phi}\left[\hat\Omega V(\phi)\right], ~~
\tilde{V}''(\tilde{\phi})=\frac{d^2\tilde{V}}{d\tilde{\phi}^2}=\tilde\Omega^{-\frac{1}{2}}\frac{d}{d\phi}\left[\tilde\Omega^{-\frac{1}{2}}\frac{d}{d \phi}\left(\hat\Omega V(\phi)\right)\right].
\end{equation*}
The scalar spectral index ($n_s$), tensor spectral index ($n_T$),
and tensor-to-scalar ratio ($r$) are the CMBR observables which are
connected to the slow-roll parameters in the following manner
\begin{equation}
n_s-1=2\tilde{\eta}_v-6\tilde{\epsilon}_v,
\end{equation}
\begin{equation}
r=16\tilde{\epsilon}_v \label{SPI}.
\end{equation}
The e-fold number $\cal N$ is defined as the ratio of the final
value of the scale factor $a_f$ during the inflationary era and
its initial value $a_i$. From its standard definition, in terms of
$H$ and $\phi$ it can be expressed as
\begin{equation}
{\cal N}=
\int_{\tilde{a_i}}^{\tilde{a_f}}\frac{d\tilde{a}}{\tilde{a}} =
\int_{\tilde{\phi_i}}^{\tilde{\phi_f}}
\frac{\tilde{H}}{d\tilde{\phi}/d\tilde{t}}d\tilde{\phi}
=\int_{\phi_i}^{\phi_f} \frac{H}{\dot{\phi}}d\phi.
\label{EFOLD}
\end{equation}
The  theoretical framework discussed above is essential for computing the inflationary parameters
in any model within the
$f(R, T)$ gravity, where a mixing term involving $R$ and $T$ is present. In the subsequent sections, we will explore two specific models characterized by a distinct inflationary potential. Our goal is to assess the viability of these models in light of current observational data.
\section{Computation of inflationary parameters for Exponential SUSY Potential}
In this section, we explore the viability of the inflation model that arises in the context of supergravity and superstrings and also in various inflations such as spin-driven inflation, brane inflation, and fiber inflation \cite{susy1, susy2, susy3, susy4}. The potential for the considered model is
\be
V(\p)=M^4 \lt(1-e^{-\l \p/\mp}\rt),
\ee
where $M$ is a constant and $\l$ is a dimensionless parameter with positive values. We have inflation for $\p/\mp > 0$. The COBE normalization \cite{cobe}, $\Delta_{R}^2=2.1\times 10^{-9}$, which is related to the amplitude of the scalar power spectrum, helps us in constraining values of free parameters. The values of the free parameters obtained from the COBE normalization are $\e=0.15\,, \x=10^8\,,\, \c=0.2\,,\,\l=1,$ and $M=.00298473$. The field value at the time of Hubble crossing $\p_f=4\mp$.
\begin{figure}[H]
\centering \subfigure[]{ \label{power_susy}
\includegraphics[width=0.48\columnwidth]{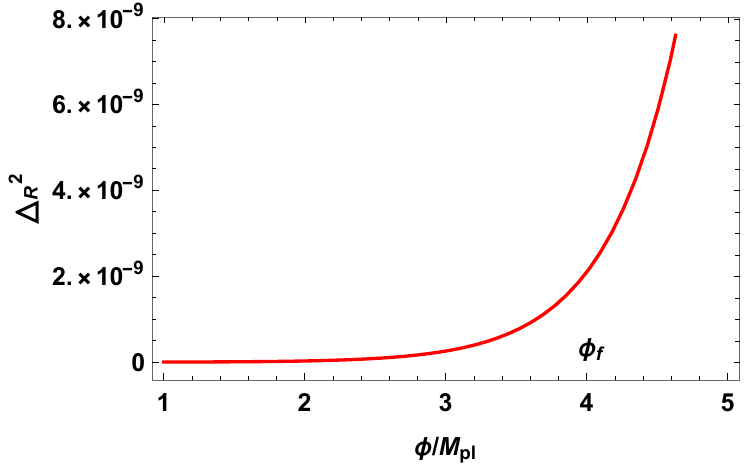}
} \subfigure[]{ \label{slow_susy}
\includegraphics[width=0.45\columnwidth]{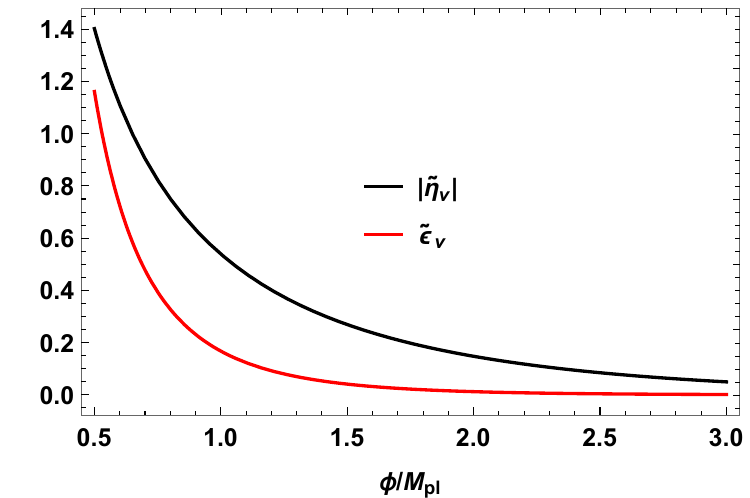}
}
\caption{The left one exhibits variation of the scalar power spectrum against $\p$, and the right one illustrates the variation of slow-roll parameters against $\p$ for the above-mentioned values of parameters.}\label{susy}
\end{figure}
Fig. (\ref{power_susy}) demonstrates the variation of the scalar power spectrum with $\p$, and Fig. (\ref{slow_susy}) shows the variation of slow-roll parameters with $\p$. At the end of inflation, the field value $\p_i$ is obtained from the condition
\be
\tep(\p_i)=1.
\ee
It is to be noted from Fig. (\ref{slow_susy}) that $\tet > 1$ at $\p_i$. This signifies the breakdown of slow-roll approximation slightly before $\p_i$, and as such, the approximation cannot describe the last few e-folds properly. Our analysis concludes the field value at the end of inflation to be $0.530288\mp$. With the requisite information, we can now obtain theoretical values of e-fold number, scalar spectral index, and tensor-to-scalar ratio. They are given by
\be
\n=51.8367\approx52,\quad  n_{s}=0.963429, \quad  \text{and} \quad  r=0.00280269.
\ee
The observed value of the spectral index is $n_s=0.9649\pm 0.0042$ at $68\%$ confidence level (CL) \cite{PLANCK} whereas for the tensor-to-scalar ratio r, at $95\%$ CL we have $r\lesssim 0.056$ from the Planck survey \cite{PLANCK} and $r \lesssim 0.036$ from BICEP/Keck array measurement \cite{bicep}.
\begin{figure}[H]
\center
\includegraphics[width=0.5\columnwidth]{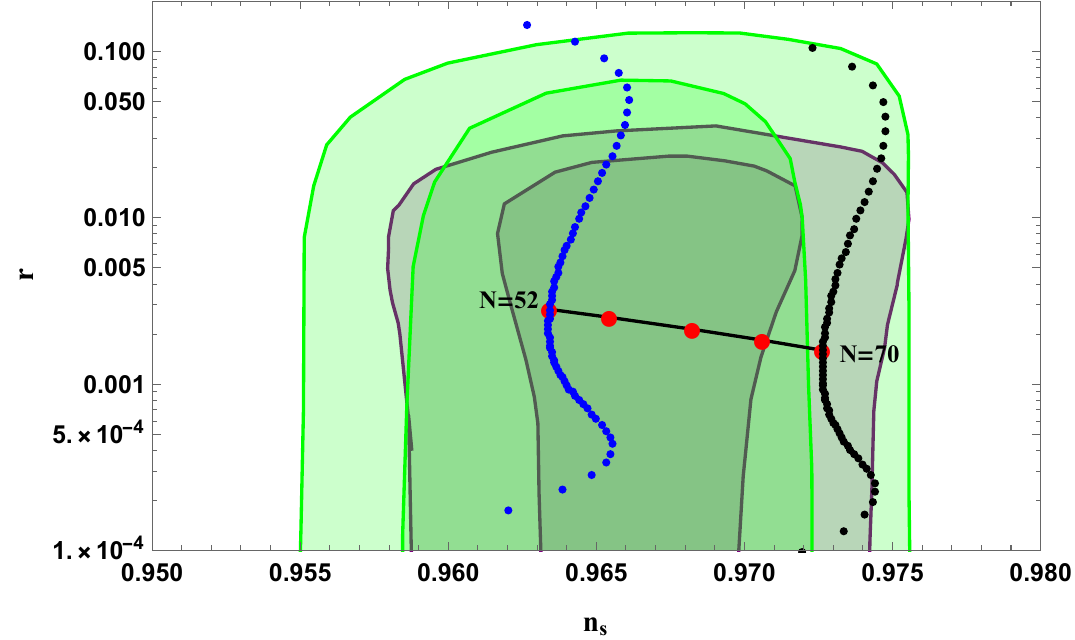}
\caption{The comparison of $n_s$ and $r$ obtained from our model with observed results from the Planck survey (Gree) and the Bicep/Keck measurement (Purple). Here, the e-fold number varies from 52 to 70. Blue and black curved lines are contour plots for fixed values of e-fold numbers mentioned in the plot. We have varied $\x$ from $-4\times 10^9$ to $4\times 10^9$. }\label{planck_susy}
\end{figure}
Thus, predicted values from our model are concordant with observed results. To further emphasize our model's efficacy, we exhibit a comparison in Fig. (\ref{planck_susy}) between theoretical and observed values where the e-fold number is varied from the obtained value, i.e., 52 to 70. Fig. (\ref{planck_susy}) clearly demonstrates the viability of our model. Fig. (\ref{planck_susy}) also displays the variation of $n_s$ and $r$ with $\x$ for fixed values of e-fold number. While $r$ always increases with $\x$, the spectral index shows the existence of two local maxima, one for a negative value of $\x$ and another for a positive value of $\x$. The values of $\x$ ($\x_c$) where $n_s$ peaks and its peak value ($n_{sc}$) both are functions of e-fold number. Our analysis finds $\x_c=-31.3532\times 10^8$ with $n_{sc}=0.96553$ and $\x_c=30.8606\times 10^8$ with $n_s=0.966102$ at $\n=52$, whereas for $\n=70$, we have $\x_c=-32.0704\times 10^8$ with $n_{sc}=0.97443$ and $\x_c=31.5864\times 10^8$ with $n_{sc}=0.974799$. $\x_c$ and $n_{sc}$ both increase with the e-fold number. Fig. (\ref{planck_susy}) clearly demonstrates the significant impact of $\x$ on observable values.

\section{Computation of inflationary parameters for a Phenomenological Potential}
In this section, We examine the commensurability of a recently introduced inflation model with observed results.
The potential for the considered model is given in \cite{pot}
\be
V(\p)=\l \mp^{4-2\a} \p^{2\a} \sin^2\lt(\frac{\b \mp^\a}{\p^\a}\rt),
\ee where $\l$ is a constant and $\a$ and $\b$ are dimensionless parameters. A similar-looking potential
can be derived from the supergravity framework \cite{pot1, pot2, pot3}. This is a modulated power-law potential, where a power-law
$\p^{2\a}$ is multiplied by a sinusoidal modulation depending on $\p^{\a}$. It combines both large-field behavior via $\p^{2\a}$ and oscillatory structure via the sinusoidal term, which can lead to interesting phenomenology like resonant particle production or modulated inflationary observables.
Proceeding as before, we utilize the COBE normalization result, which yields the following values of various parameters
\be
\e=0.001,~ \x=10^9,~ \c=0.0001,~ \a=2.5,~ \b=2.0,~ \text{and} ~~ \l=5.18314\times 10^{-12}.\label{val_new}
\ee
The field value at the time of horizon exit or Hubble crossing is found to be $\p_{f}=3.0 \mp$. Inflation ends when we have $\tep(\p_i)=1$ where $\p_i$ is the field value at the end of inflation. We explore the variation of the scalar power spectrum $\Delta_{R}^2$ and the slow-roll parameters $\tep$ and $\tet$ in Fig. (\ref{new}).
\begin{figure}[H]
\centering \subfigure[]{ \label{power_new}
\includegraphics[width=0.5\columnwidth]{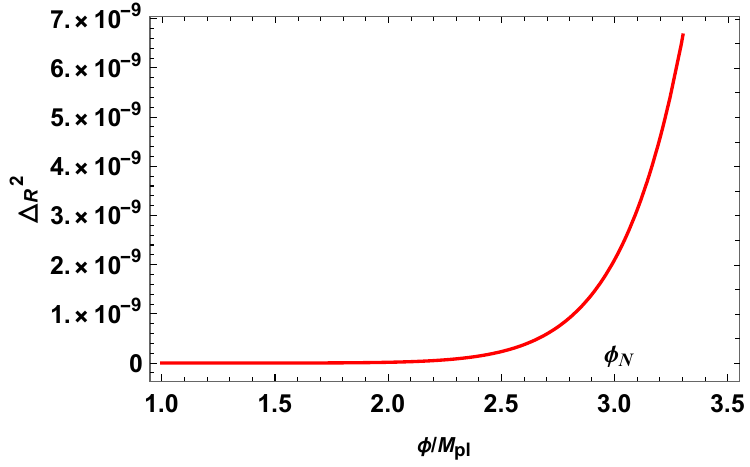}
} \subfigure[]{ \label{slow_new}
\includegraphics[width=0.45\columnwidth]{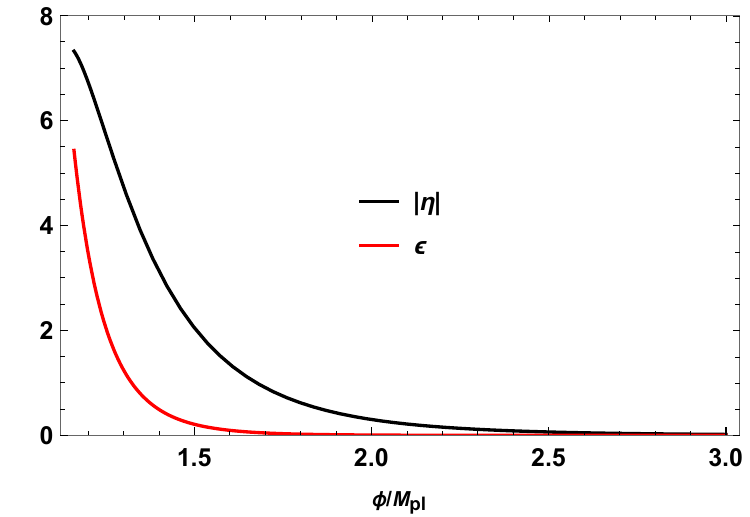}
}
\caption{The left one exhibits variation of the scalar power spectrum against $\p$, and the right one illustrates the variation of slow-roll parameters against $\p$ for the above-mentioned values of parameters.}\label{new}
\end{figure}
Similar to the SUSY potential, here, too, we see $\tet(\p_i)>1$, signifying the breakdown of slow-roll approximation, and hence, the last few e-folds cannot be described by the approximation. Our analysis finds $\p_i=1.32279 \mp$.
\begin{figure}[H]
\center
\includegraphics[width=0.45\columnwidth]{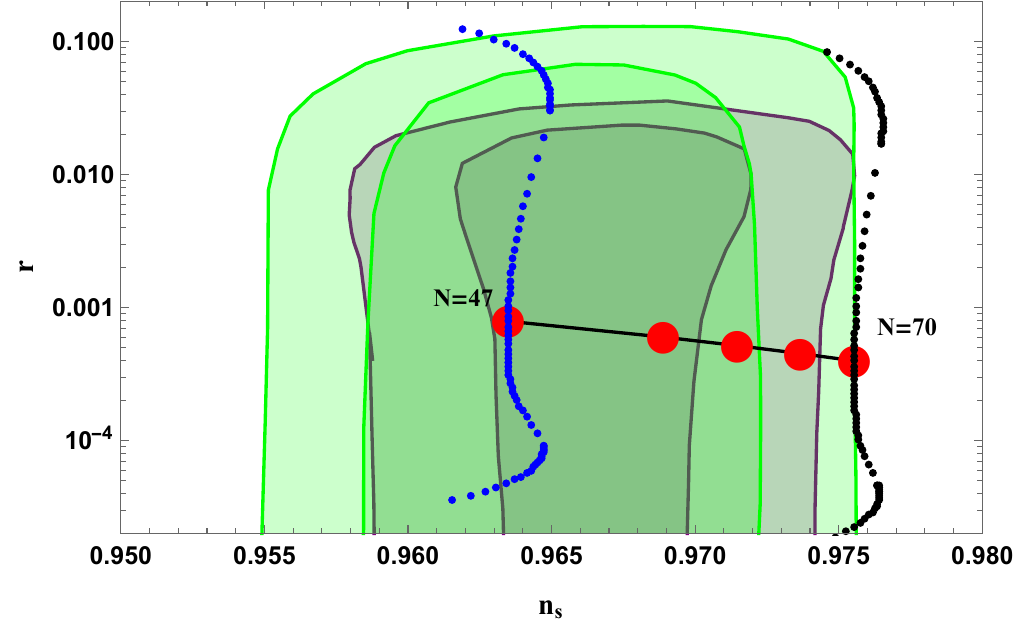}
\caption{The comparison of $n_s$ and $r$ obtained from our model with observed results from the Planck survey (Gree) and the Bicep/Keck measurement (Purple). Here, the e-fold number varies from 47 to 70. Blue and black curved lines are contour plots for fixed values of e-fold numbers mentioned in the plot. We have varied $\x$ from $-12\times 10^9$ to $12\times 10^9$. }\label{planck_new}
\end{figure}
With all the necessary information required to obtain the e-fold number, spectral index, and tensor-to-scalar ratio at our disposal, we can now examine whether our model is concordant with observed results or not. Predicted values of  e-fold number, spectral index, and the tensor-to-scalar ratio from our model are
\be
\n=46.8373\approx 47, \quad n_s=0.963477, \quad \text{and} \quad r=0.000786256.
\ee

Fig. (\ref{planck_new}) compares our predicted values with observed data from the Planck 2018 survey \cite{PLANCK} and BICEP/Keck array measurement \cite{bicep}. Our results are in excellent agreement with observed values, positioning the proposed model as a strong candidate for explaining inflation. Fig. (\ref{planck_new}) also illustrates the variation of $n_s$ and $r$ with $\x$ for fixed values of e-fold number. Similar to the exponential SUSY inflation model, we also observe two local maxima of $n_s$ for fixed e-fold numbers. We obtain $\x_c=-11.127\times 10^9$ with $n_{sc}=0.964709$ and $\x_c=11.1573\times 10^9$ with $n_{sc}=0.964953$ at $\n=47$ while at $\n=70$, we have $\x_c=-11.2754\times 10^9$ with $n_{sc}=0.976417$ and $\x_c=11.2912\times 10^9$ with $n_{sc}=0.97653$.
\section{Conclusion}
In this manuscript, we consider the $f(R, T)$ gravity where, apart from the $R$ term, we have conformal rescaling and RT-mixing terms. The RT-mixing term introduces non-minimal matter-gravity coupling that exhibits important implications in astrophysical and cosmological scales. Observed results from the Planck 2018 survey and BICEP/Keck array measurement are potent tools to test different inflation models. Against the backdrop of this $f(R, T)$ gravity framework, we put two inflation models against observed results from the Planck 2018 survey and BICEP/Keck array measurement to test. We utilized the COBE normalization value to constrain the different parameters involved and extract the field value at the time of Hubble crossing.

The first inflation model we explored against the backdrop of $f(R, T)$ gravity is the exponential SUSY model, which arises in supergravity and superstrings and appears in spin-driven, brane, and fiber inflations. With the help of COBE normalization, we first constrained values of different parameters involved, such as $\e,~ \x,~ \c,~\l$, and $M$, and the field value at the Hubble crossing. The condition $\tep(\p_i)=1$ provides the field value $\p_i$ at the end of inflation. Finally, we calculate the theoretical values of the spectral index and the tensor-to-scalar ratio, which come out to be $0.963239$ and $0.00266874$, respectively. These values are consistent with Planck and BICEP/Keck data, as shown in Fig. (\ref{planck_susy}).

Having proven the viability of the SUSY model in the considered $f(R, T)$ theory, we examine a novel inflation model that has recently been introduced. The model shows promising features in the evolution of post-inflation. Similar to the previous case, we extricate values of different parameters from COBE normalization, and the field value at the end of inflation is extracted from the condition $\tep(\p_i)=1$. Then, the theoretical values of observables $n_s$ and $r$ are obtained, which came out to be $0.96323$ and $0.00067115$. These values are completely concordant with observed results. The excellent agreement between theoretical predictions and observed results is emphasized graphically in Fig. (\ref{planck_new}).

We also examine the variation of $n_s$ and $r$ against $\x$ for fixed values of e-fold number and display them in Fig. (\ref{planck_susy}) and (\ref{planck_new}). In both cases, the spectral index possesses two local maxima, one for a negative value of $\x$ and another for a positive value $\x$. We have reported values of $\x_c$ and $n_{sc}$ for selective values of e-fold number for both inflationary models. The impact of the RT-mixing term on observables is found to be significant. We conclude this section by reiterating our observations in Sec. (III) and (IV) that the slow-roll approximation, which provides the backbone of our analysis, breaks down near the end of inflation. At the end of inflation, we can no longer neglect the time derivatives of the inflation, and as such, the last few e-folds cannot be adequately described by the slow-roll approximation.

\textbf{Acknowledgements:} SKJ would like to thank Li-Yang Chen, Hongwei Yu, and Puxun Wu for their help in this study.

\end{document}